\documentclass[aps,pra,superscriptaddress,twocolumn]{revtex4-2}
\usepackage{dcolumn}
\usepackage{amsmath}
\usepackage{amsfonts}
\usepackage{amssymb}
\usepackage{graphicx}
\usepackage{hyperref}
\usepackage{xcolor}
\usepackage{ulem}
\usepackage{bookmark}

\begin{document}
\title{Experimental Investigation of Twist Conservation in Nonlinear Optical Three-Wave Mixing}

\author{G. H. dos Santos}
\email{fisica.gu@gmail.com}
\affiliation{Departamento de F\'{i}sica, Universidade Federal de Santa Catarina, CEP 88040-900, Florian\'{o}plis, SC, Brazil}
\affiliation{Departamento de F\'{\i}sica, Universidad de Concepci\'on, 160-C Concepci\'on, Chile}
\affiliation{Millennium Institute for Research in Optics, Universidad de Concepci\'on, 160-C Concepci\'on, Chile}

\author{A. L. S. Santos Junior }
\affiliation{Instituto de F\'{i}sica, Universidade Federal Fluminense, 24210-346 Niter\'{o}i, RJ, Brazil}

\author{M. Gil de Oliveira}
\affiliation{Instituto de F\'{i}sica, Universidade Federal Fluminense, 24210-346 Niter\'{o}i, RJ, Brazil}

\author{A. C. Barbosa}
\affiliation{Instituto de F\'{i}sica, Universidade Federal Fluminense, 24210-346 Niter\'{o}i, RJ, Brazil}
\author{B. Pinheiro da Silva}
\affiliation{Instituto de F\'{i}sica, Universidade Federal Fluminense, 24210-346 Niter\'{o}i, RJ, Brazil}

\author{N. Rubiano da Silva}
\affiliation{Departamento de F\'{i}sica, Universidade Federal de Santa Catarina, CEP 88040-900, Florian\'{o}plis, SC, Brazil}

\author{G.~Ca\~{n}as}
\affiliation{Departamento de F\'isica, Universidad del B\'io-B\'io, Collao 1202, 5-C Concepci\'on, Chile}

\author{S. P. Walborn}
\email{swalborn@udec.cl}
\affiliation{Departamento de F\'{\i}sica, Universidad de Concepci\'on, 160-C Concepci\'on, Chile}
\affiliation{Millennium Institute for Research in Optics, Universidad de Concepci\'on, 160-C Concepci\'on, Chile}

\author{A. Z. Khoury}
\affiliation{Instituto de F\'{i}sica, Universidade Federal Fluminense, 24210-346 Niter\'{o}i, RJ, Brazil}

\author{P. H. Souto Ribeiro}
\email{p.h.s.ribeiro@ufsc.br}
\affiliation{Departamento de F\'{i}sica, Universidade Federal de Santa Catarina, CEP 88040-900, Florian\'{o}plis, SC, Brazil}

 \begin{abstract}
We conduct an experimental investigation into the conservation of the twist phase in Twisted Gaussian Schell Model (TGSM) beams during both up- and down-conversion three-wave mixing nonlinear processes. Independently generated TGSM beams, prepared with varying twist parameters, are used to pump and seed the nonlinear interactions. The resulting beams are then analyzed to determine their twist properties.
Our findings demonstrate that the twists of the up- and down-converted beams depend on those of the pump and seed beams. Additionally, the results indicate that the twist phase is conserved throughout the process, in qualitative agreement with theoretical predictions. This study is motivated by the increasing potential applications of TGSM beams in various fields.

 \end{abstract}

\maketitle

\section{Introduction}

Since the invention of the laser, coherent optical fields have been the ones most extensively studied and widely applied in science and technology. However, an important class of partially coherent optical fields is now gaining increasing attention—not only for exploring fundamental properties of light but also for their promising practical applications.

The Gaussian Schell Model (GSM) was introduced by A. C. Schell in 1961 \cite{allan61}. This model describes partially coherent stochastic optical fields, which are characterized by their transverse coherence length and exhibit  propagation properties that are distinct from their coherent counterparts.

In 1993, Simon and Mukunda proposed a generalization of the Gaussian Schell Model through the introduction of a twist phase, thereby establishing the Twisted Gaussian Schell Model (TGSM) beams \cite{simon98}. This extended formulation incorporated an additional parameter to quantify the twist phase. Unlike naturally occurring optical fields, TGSM beams require artificial synthesis, with current laboratory methods producing only approximations of ideal TGSM beams \cite{Cai17,wang20,tian20}.

These beams exhibit remarkable propagation characteristics and exceptional resistance to phase perturbations, making them particularly valuable for optical communication systems \cite{wang10,liu19,zhou20}. From a fundamental research standpoint, TGSM beams demonstrate intriguing properties for generating photonic entangled states \cite{Hutter20,Hutter21,ismail2020} and engaging in nonlinear optical interactions \cite{Santos2022}.

The twist phase is fundamentally connected to the orbital angular momentum (OAM) of light, though this relationship presents several counterintuitive aspects. Notably, the twist phase increases as the transverse coherence length decreases—a surprising behavior given that OAM typically requires fully coherent wavefronts with a linear phase dependence on the azimuthal angle in the transverse plane.

 For coherent OAM-carrying beams, well-defined conservation laws govern their behavior in nonlinear optical processes such as parametric up- and down-conversion in bulk crystals \cite{mair01,caetano02,walborn04a}. Moreover, besides the simple conservation of topological charge in these processes, there are surprising consequences related to the mode structure of the interacting beams, leading to mode selection rules \cite{Pereira17,Oliveira21}. However, the extension of these conservation principles to twisted partially coherent beams has not yet been studied in detail. 
  
In this work, we investigate the conservation of the twist phase in Twisted Gaussian Schell Model (TGSM) beams undergoing nonlinear interactions via up- and down-conversion processes in bulk nonlinear crystals. Specifically, we experimentally examine second harmonic generation (SHG) pumped by TGSM beams, as well as stimulated parametric down-conversion (StimPDC) processes where TGSM beams serve as both pump and seed. By systematically varying both the transverse coherence length and twist parameter of the input TGSM beams, we characterize the second harmonic and stimulated output beams to evaluate twist conservation in these nonlinear processes. Our results demonstrate, for the first time, experimental evidence of qualitative twist conservation in both SHG and StimPDC interactions.

 \section{Theory}
 
\begin{figure*}
      \centering
      \includegraphics[width=\linewidth]{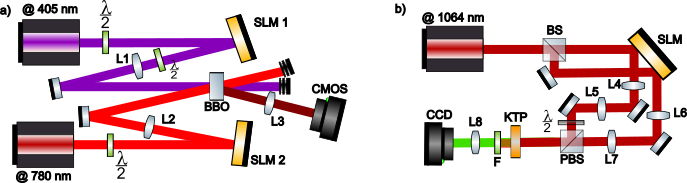}
      \caption{Experimental setups for measuring the far-field transverse profile of the converted beams: a) the idler beam in the stimulated parametric down-conversion. b) the second harmonic beam in the up-conversion process.}
      \label{fig:setup}
\end{figure*}

\subsection{Stimulated Parametric Down-conversion StimPDC}

In Ref. \cite{Santos2022}, the cross-spectral density (CSD) is derived for stimulated down-conversion (StimPDC) and the relationship between the twist parameters of pump, signal and idler beams is obtained. In StimPDC,  a pump beam, usually in the blue-violet wavelength range, is incident upon a nonlinear crystal. It produces spontaneous emission of pairs of photons, called signal ($s$) and idler ($i$), whose wavelengths are in the  red to near-infrared range. A second seed laser is aligned with the signal beam and stimulates the emission in its wavelength. As a result, the emission in the idler conjugate beam is also stimulated due to the fact that the signal and idler are produced as a pair. Here, we are interested in the properties of the idler beam and its dependence on the pump and seed fields. For coherent seed and pump fields, the idler field can be written as 
\begin{equation}
\label{SP}
    \mathcal{E}_{SP}(\mathbf{r}) = i \, g_{SP} \, \mathcal{E}_{p}(\mathbf{r}) \, \mathcal{E}^{\ast}_{s}(\mathbf{r}),
\end{equation}
where $\mathcal{E}_{SP}(\mathbf{r})$ is the idler field, $\mathcal{E}_{p}(\mathbf{r})$ is the pump field, $\mathcal{E}_{s}(\mathbf{r})$ is the seed field, and $g_{SP}$ is the effective StimPDC coupling constant. We assume the stationary case, in which the temporal degrees of freedom are not relevant, and that the spontaneous emission is negligible in comparison to the stimulated one. 

Let us assume now that pump and seed are perfect TGSM beams. Pumping and seeding the process with uncorrelated TGSM beams allows easy calculation of the CSD of the idler beam: 
\begin{equation}
\label{CSDidler}
    W_{SP}(\mathbf{r},\mathbf{r'}) = g_{SP}^2 \, W_{p}(\mathbf{r},\mathbf{r'}) \, W_{s}^{\ast}(\mathbf{r},\mathbf{r'}),
\end{equation}
where the CSD is defined as:

\begin{equation}
\label{CSD}
    W(\mathbf{r},\mathbf{r}^\prime) = \langle \mathcal{E}(\mathbf{r}) \mathcal{E}^*(\mathbf{r}^\prime) \rangle,
\end{equation}
and $\mathbf{r} = (x,y)$ is the transverse position and $\langle \cdot \rangle$ denotes the ensemble average.

TGSM beams are stochastic fields characterized by the CSD:
\begin{eqnarray}
\label{TGSM}
    W_{TGSM}(\mathbf{r},\mathbf{r^{\prime}}) &=& A \, e^{-\frac{r^2 + r^{\prime2}}{4w^2}}
  e^{-\frac{|\mathbf{r} - \mathbf{r^{\prime}}|^2}{2\delta^2}} e^{- i k \frac{|\mathbf{r}^2 - \mathbf{r^{\prime 2}}|}{2R}}  \\ \nonumber
  &\times&  e^{-i k \mu (x y^{\prime} - y x^{\prime})},
\end{eqnarray}
where $A$ is constant, $w$ is the beam waist, $\delta$ is the transverse coherence length, $k$ is the wavenumber, $R$ is the  radius of curvature, and $\mu$ is the twist phase.   

Returning to Eq. \eqref{CSDidler}, the CSD for the idler output field in StimPDC is given by the product of the CSD of the pump beam and the complex conjugate of the CSD of the seed beam. We can write explicitly the CSD for the generated idler beam \cite{Santos2022}: 

\begin{eqnarray}
\label{WSP}
    W_{SP}(\mathbf{r},\mathbf{r^{\prime}}) &=& A_i \, e^{-\frac{r^2 + r^{\prime2}}{4w_i^2}}
  e^{-\frac{|\mathbf{r} - \mathbf{r^{\prime}}|^2}{2\delta_i^2}} e^{- i k_i \frac{|\mathbf{r}^2 - \mathbf{r^{\prime 2}}|}{2R_i}}  \\ \nonumber
  &\times&  e^{-i k_i \mu_i (x y^{\prime} - y x^{\prime})},
\end{eqnarray}
where $A_i$ is a constant,

\begin{equation}
w_i^2= \frac{w_s^2 w_p^2}{w_s^2 + w_p^2},
\label{eq:wid}
\end{equation}

\begin{equation}
\delta_i^2= \frac{\delta_s^2 \delta_p^2}{\delta_s^2 + \delta_p^2},
\label{eq:deltaid}
\end{equation}

\begin{equation}
\frac{k_i}{R_i}= \frac{k_p}{R_p}-\frac{k_s}{R_s},
\label{eq:Rid}
\end{equation}
and
\begin{equation}
 k_i \mu_i = k_p \mu_p - k_s \mu_s,
\label{eq:tpd}
\end{equation} 
recalling that $i,s,p$ stand for idler, seed and pump, respectively.  Comparing with the definition of TGSM beams \eqref{TGSM}, we conclude that when 
pump and seed are independent TGSM beams, the generated idler beam is also a TGSM beam whose
parameters are related to those of the pump and seed by Eqs. \eqref{eq:wid}, \eqref{eq:deltaid}, \eqref{eq:Rid}, and \eqref{eq:tpd}.  By combining Eqs. \eqref{eq:deltaid} and \eqref{eq:tpd}, one can find a simple relation between the normalized twist phases $\tau=\mu k \delta^2$ of the interacting beams:
\begin{equation}
    \tau_i = \frac{\delta_s^2}{\delta_s^2 + \delta_p^2}\,\tau_p - \frac{\delta_p^2}{\delta_s^2 + \delta_p^2}\,\tau_s.
    \label{eq:taui}
\end{equation}

It is important to notice that Eq. \eqref{eq:tpd} expresses the twist conservation for StimPDC with uncorrelated pump and seed. Moreover, Eq. \eqref{eq:taui} indicates that the idler normalized twist phase depends on both input twist phases, weighted by the relative coherence lenghts.

\subsection{Second Harmonic Generation SHG}

In the up conversion (UC) process, the second harmonic (SH) field generated by two input pumping beams can be written as 
\begin{equation}
\label{SH}
    \mathcal{E}_{SH}(\mathbf{r}) = i \, g_{UC} \, \mathcal{E}_{p1}(\mathbf{r}) \, \mathcal{E}_{p2}(\mathbf{r}),
\end{equation}
where $g_{UC}$ is the effective coupling constant and $1,2$ refer to the pumping beams that can be provided by either only one beam or two independent beams. Again, for simplicity, we are not considering the temporal degrees of freedom, as we are interested only in the transverse spatial degrees of freedom. Eqs. \eqref{SP} and \eqref{SH} are similar and the difference between them is that there is no complex conjugation of either of the pumping beams for the UC process. 

Pumping the process with two uncorrelated TGSM beams allows us to easily calculate the CSD of the output (SH) beam: 
\begin{equation}
\label{CSD1}
    W_{SH}(\mathbf{r},\mathbf{r'}) = g_{UC}^2 \, W_{p1}(\mathbf{r},\mathbf{r'}) \, W_{p2}(\mathbf{r},\mathbf{r'}),
\end{equation}
which can be written in terms of the parameters of the input fields as: 

\begin{eqnarray}
\label{WSP}
    W_{SH}(\mathbf{r},\mathbf{r^{\prime}}) &=& A_{sh} \, e^{-\frac{r^2 + r^{\prime2}}{4w_{sh}^2}}
  e^{-\frac{|\mathbf{r} - \mathbf{r^{\prime}}|^2}{2\delta_{sh}^2}} e^{- i k_{sh} \frac{|\mathbf{r}^2 - \mathbf{r^{\prime 2}}|}{2R_{sh}}}  \\ \nonumber
  &\times&  e^{-i k_{sh} \mu_{sh} (x y^{\prime} - y x^{\prime})},
\end{eqnarray}
where $A_{sh}$ is a constant, and

\begin{equation}
w_{sh}^2= \frac{w_{p1}^2 w_{p2}^2}{w_{p1}^2 + w_{p2}^2},
\label{eq:wi}
\end{equation}

\begin{equation}
\delta_{sh}^2= \frac{\delta_{p1}^2 \delta_{p2}^2}{\delta_{p1}^2 + \delta_{p2}^2},
\label{eq:deltai}
\end{equation}

\begin{equation}
\frac{k_{sh}}{R_{sh}}= \frac{k_{p1}}{R_{p1}} + \frac{k_{p2}}{R_{p2}},
\label{eq:Ri}
\end{equation}
and
\begin{equation}
 k_{sh} \mu_{sh} = k_{p1} \mu_{p1} + k_{p2} \mu_{p2}.
\label{eq:tp1}
\end{equation} 

In conclusion, the CSD for the output field in the up-conversion is given by the product of the CSDs of the two pumping and independent fields, which can be written in the form of a TGSM beam. From the relationship between input and output fields, we get the conservation law for the twist given by Eq. \eqref{eq:tp1}. In analogy to what we showed for StimPDC, the normalized twist phase of the output beam is a sum of the input twist phases, weighted by their relative coherence lengths:
\begin{equation}
    \tau_{sh} = \frac{\delta_{p2}^2}{\delta_{p1}^2 + \delta_{p2}^2}\,\tau_{p1} + \frac{\delta_{p1}^2}{\delta_{p1}^2 + \delta_{p2}^2}\,\tau_{p2}.
    \label{eq:taush}
\end{equation}

\section{Experiment}

 \begin{figure*}
 \includegraphics[width=\linewidth]{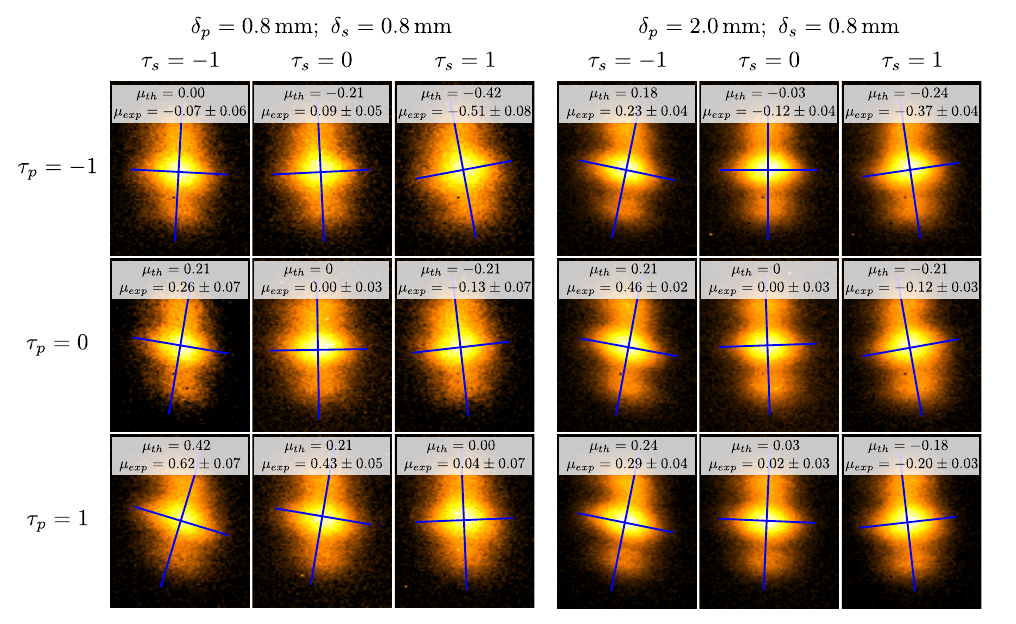}
 \caption{Double-slit interference patterns of the idler beam generated by different combinations of twist of pump and seed beams. Each line represents the same value of normalized twist of the pump beam, being -1, 0, and 1 from top to bottom. Likewise, each column corresponds, from left to right, to the values of -1, 0, and 1 for the normalized twist of the seed beam. The theoretical and experimental values of twist phase of the idler, in $10^{-3}$~mm$^{-1}$, are shown as insets.}
 \label{fig:QaulitativeTwist}
 \end{figure*}

  \begin{figure*}
 \includegraphics[width=\linewidth]{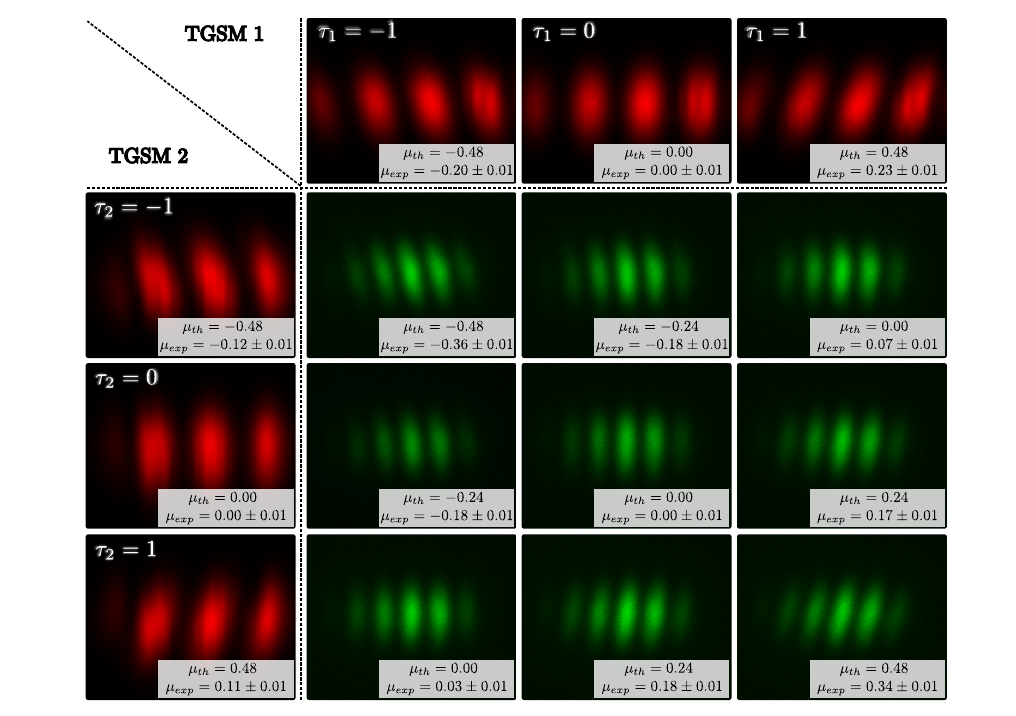}
 \caption{Double-slit interference patterns of the second harmonic generated by different combinations of TGSM pump beams. From top to bottom, each line corresponds to a value of normalized twist of the pump 2 beam of -1, 0, and 1. Likewise, each column represents the values of -1, 0, and 1 for the normalized twist of the pump 1 beam, from left to right. The theoretical and experimental values of twist phase of all beams, in $10^{-3}$~mm$^{-1}$, are shown as insets.}
 \label{fig:SHG}
 \end{figure*}

We performed two experiments to investigate the conservation of twist in nonlinear three-wave mixing optical processes: StimPDC and SHG.
  
\subsection{StimPDC}

The StimPDC experimental setup is sketched in Fig. \ref{fig:setup}a. A continuous-wave laser diode oscillating at 405 nm is prepared as a TGSM beam by means of the spatial light modulator SLM1 (Holoeye Leto 3), which displays a dynamical set of phase masks consisting in the phase randomized coherent field method \cite{tian20,canas22}. Lens L1 is used to image the SLM1 plane onto the BBO nonlinear crystal. BBO stands for beta barium borate, and the crystal is cut for a type-I phase-matching interaction. A half waveplate $\lambda/2$  is used to ensure that the blue laser is horizontally polarized at the surface of SLM1. A second half waveplate $\lambda/2$ is used to rotate the polarization of the blue TGSM beam emerging from SLM1 to the vertical direction, which is required to pump the BBO crystal. 

The seed beam is also prepared as a TGSM beam by means of the same method just described.  A 780 nm laser diode is spatially modulated at SLM2 (Holoeye Leto 3). Lens L2 images the SLM2 surface onto the BBO crystal plane so that it interacts with the blue TGSM beam via the nonlinear medium. Phase-matching conditions are achieved by proper alignment of the pump and seed beams with respect to the incidence plane and the optical axis of the BBO crystal. As a result of the nonlinear optical interaction, an idler beam is generated. Lens L3 ($f=400$~mm) performs the optical Fourier Transform of the idler beam having the crystal plane as the input plane and the sCMOS camera (Thorlabs CS2100M) as the output plane. 

In order to analyze the transfer of twist from the pump and seed TGSM beams to the generated idler beam, we use a method based on double-slit interference \cite{canas22}. The slits are imprinted in the idler beam by means of a modulation in the seed beam. This is equivalent of seeding the nonlinear interaction with a light beam that looks like a TGSM beam that crosses a double-slit. The result is a double-slit pattern of the idler beam in the far field. The interference patterns allow us to make a qualitative analysis of the twist.

\subsection{SHG}

The SHG experimental setup is sketched in Fig. \ref{fig:setup}b. A continuous-wave, horizontally polarized diode laser at 1064 nm wavelength is split in two beams using a 50/50 beam splitter. Both beams are directed to an SLM so that each half of it modulates one of the beams independently. We use the phase randomized coherent field method \cite{tian20,canas22} to generate two independent TGSM beams with different parameters. After modulation, the polarization of one of the beams is changed to vertical using a half-wave plate $\lambda/2$. 
The two beams are recombined with the aid of a polarizing beam splitter (PBS) so that we have two collinear, independent TGSM beams with orthogonal polarizations and independent controllable parameters. They are sent to a potassium titanyl phosphate (KTP) type-II phase-matching crystal for up-conversion nonlinear interaction. Lenses L4, L5, L6 and L7 are used to image the SLM plane onto the KTP crystal plane. The second harmonic generated field at 532 nm is filtered out from the fundamental beams and measured in the far-field by means of optical Fourier transform with lens L8 ($f=$ 200 mm). The intensity patterns are registered with a CCD camera.


\section{Results}

\subsection{StimPDC}
The results for StimPDC are presented in Fig. \ref{fig:QaulitativeTwist}. All panels display interference patterns of double-slit diffraction for the idler beam resulting from pumping and seeding the StimPDC with TGSM beams. The goal is to relate the pump and seed twists with the idler beam twist and interpret it in terms of twist conservation in the process. Each large panel shows the interference patterns for different combinations of twists of pump and seed beams with corresponding transverse coherence lengths of pump $\delta_p$ and seed $\delta_s$ (shown on the top). In the right-hand panel, $\delta_p$ is increased compared to the left-hand one.
We use the blue lines as guides for the eye, indicating the rotation of the patterns about the axis perpendicular to the plane of the images. They provide a qualitative manifestation of the predicted twist conservation. 

A quantitative analysis is performed starting with a nonlinear curve fitting of the 2D interference patterns. The fitting function is an asymmetric 2D Gaussian distribution modulated by a displaced 1D cosine function along its main axis. The main axis is rotated by the tilt angle $\phi$ about the axis orthogonal to the image plane. From $\phi$, we compute the twist phase as $\mu=\mathrm{atan} (\phi)/f$ \cite{Santos2022}. The results are shown in the insets in Fig. \ref{fig:QaulitativeTwist} accompanied by the respective expected value given by $\mu=\tau/(k\delta^2)$, with $\tau_i$ from Eq. \eqref{eq:taui}. 

Concerning conservation of twist phase in StimPDC, our experimental results agree qualitatively with the theoretical predictions. However, strong quantitative agreement is lacking for most of the cases analyzed. The reason for this discrepancy is most likely due to the phase-randomized coherent field method \cite{tian20,canas22} used to generate the beams. This technique consists in displaying a sequence of holograms on the SLM, and indeed it was shown in Ref. \cite{canas22} that the residual  coherence background that is present in the TGSM beam can have a large impact on coherence measurements of the field.  We note that this effect is present in some extent in all beams, for both experiments.  Moreover, the  nonlinear processes (both StimPDC and SHG) occur on a time scale that is much smaller than the integration time required to obtain a good approximation of a TGSM beam. Therefore, on the time scale of the nonlinear processes, these are not true TGSM beams interacting inside the crystals.  We believe this to be an essential issue, and it will be resumed in the discussion section below. 

\subsection{SHG}

The results for SHG are presented in Fig. \ref{fig:SHG}. All panels display interference patterns of double-slit diffraction for the second-harmonic beam resulting from two fundamental TGSM beams. The goal is once more to relate the twist phase of the fundamental pump beams with the twist phase of the SHG beam and interpret it in terms of twist conservation in the SHG process. The central 9$\times$9 panel shows the interference patterns for different combinations of twists of the pump beams. The normalized twist phases of the pump beams $\tau_{p1}$ and $\tau_{p2}$ are shown on the left and top panels, respectively. These results are qualitatively consistent with the conservation law. We note that both pumps have the same coherence length and thus the twist conservation expressed in Eq.~\ref{eq:taush} can be easily rewritten as $\tau_{sh} = 0.5(\tau_{p1} + \tau_{p2})$. 

A quantitative analysis was also performed by obtaining rotation angles after a nonlinear curve fit of the 2D interference patterns, using the same procedure as was done for StimPDC. The fit function is similar, except that in this case there is no need to consider the shape of the emission cone, because the fundamental and SHG beams are all collinear and beam-like. The results are shown in the insets in Fig. \ref{fig:SHG}, accompanied by the respective theoretical predictions obtained from Eq. \ref{eq:taush}. Again, a strong quantitative agreement is lacking, most likely for the reasons discussed above.

\section{Discussion}

To study GSM and TGSM beams and generate them with controllable parameters—such as transverse coherence length and twist—we must synthesize them in the laboratory. To our knowledge, there are no natural sources of TGSM light beams, and while GSM beams do occur naturally, their properties cannot be readily controlled. Additionally, natural GSM sources are weak, exhibiting low photon number occupation per mode, and thus unsuitable to pump nonlinear optical processes.

\begin{figure*}
 \includegraphics[width=\linewidth]{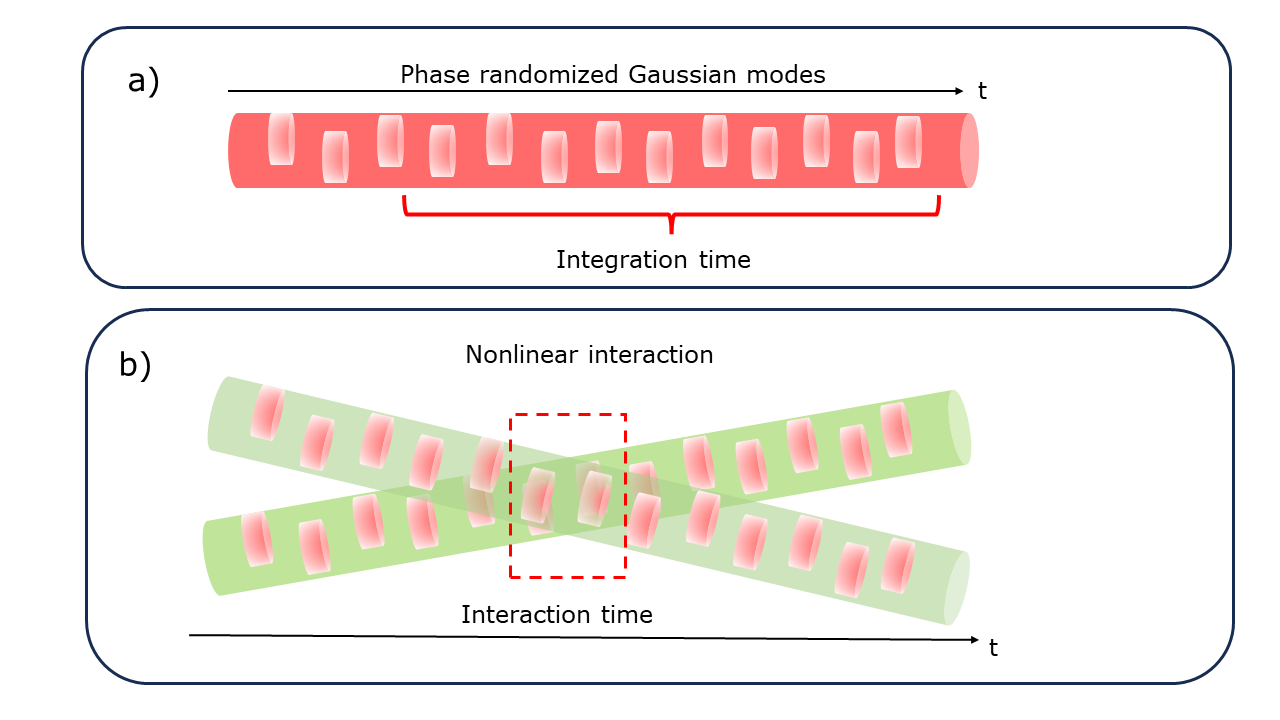}
 \caption{A TGSM beam generated by the phase-randomized-Gaussian mode method is represented schematically in the upper part of the figure. It is composed by a temporal sequence of displaced Gaussian modes having no phase correlation with each other. In the lower part of the figure, two TGSM beams are schematically represented interacting within a nonlinear medium. The integration time required to recover the TGSM properties is much longer than the interaction time of the nonlinear process.}
 \label{fig:interaction}
 \end{figure*}

From the perspective of generating intense TGSM beams with highly controllable parameters, methods based on sequences of coherent modes present an excellent solution. For this reason, our work employs an approach that generates sequences of Gaussian modes displaced from the propagation axis, each with random relative phases. Figure \ref{fig:interaction} illustrates both the longitudinal (a) structure of a synthesized TGSM beam and the nonlinear interaction between two such beams in a medium (b).
A key consideration is the need for time integration over multiple phase-randomized coherent modes to achieve a TGSM beam with the desired parameters. Since the nonlinear interaction occurs almost instantaneously, the output beam consists of a sequence of modes resulting from the interaction between two input coherent modes.  
It is necessary that the integration time together with the sequence of independent SLM holograms is sufficient to guarantee that the ensemble average of the product of the fields is equivalent to the product of the averages. To best  (but not completely) achieve this with the phase-randomized method, it would be necessary to guarantee that all $N$ holograms used to generate one TGSM input beam are paired with all $M$ holograms used to generate the other.  Since it requires on the order of $N\sim M \sim 300$ to reduce residual coherence to a few \%, we would require a total of $10^5$ hologram pairs, which, given the low intensity of the output beams and slow refresh rate of the SLMs, would require extremely long data acquisition times (at least several hours),  at which point other important imperfections would become relevant, such as beam drift. 

We note that it has been shown that correlations between the phase-randomization of the input TGSM beams can indeed lead to an output beam (from a nonlinear process) that no longer conforms to the TGSM one \cite{deOliveira24}. In the present case, we do indeed find that the generated SHG and StimPDC fields  qualitatively adhere to a twist conservation law. However, due to the aforementioned effects, the theoretical description of this conservation, which assumes stationary fields, does not fully hold.

 \section{Conclusion}
 
We present a theoretical and experimental investigation of a conservation law governing the twist of Twisted Gaussian Schell Model (TGSM) beams undergoing nonlinear interactions. Two experiments were conducted: one employing TGSM beams as pumps for second-harmonic generation (SHG), and another utilizing TGSM beams as input optical fields in stimulated parametric down-conversion (StimPDC). The experimental results agree with theoretical predictions, supporting the conservation of twist phase in the nonlinear processes. However, while qualitative agreement is observed, quantitative discrepancies remain significant. We attribute this to limitations in the TGSM beam synthesis method. Our findings highlight the need for improved techniques to achieve better control over intense TGSM beams for use in nonlinear optical processes.  

%



%


 \begin{acknowledgements}
This work has been supported by the following Brazilian research agencies Conselho Nacional de Desenvolvimento Cient\'{\i}fico e Tecnol\'ogico (CNPq - DOI 501100003593), Coordena\c c\~{a}o de Aperfei\c coamento de Pessoal de N\'\i vel Superior (CAPES DOI 501100002322), Funda\c c\~{a}o de Amparo \`{a} Pesquisa do Estado de Santa Catarina (FAPESC - DOI 501100005667), Instituto Nacional de Ci\^encia e Tecnologia de Informa\c c\~ao Qu\^antica (INCT-IQ 465469/2014-0 e INCT-IQNano 406636/2022-2); the Chilean agencies Fondo Nacional de Desarrollo Cient\'{i}fico y Tecnol\'{o}gico (FONDECYT - DOI 501100002850) (1230796, 1240746); National Agency of Research and Development (ANID) Millennium Science Initiative Program—ICN17sr\_h012. 
 \end{acknowledgements}

 \bibliographystyle{apsrev}
 \bibliographystyle{unsrt}
 \bibliography{main}

\begin{thebibliography}{10}

\bibitem{allan61}
Allan~Carter Schell.
\newblock {\em The multiple plate antena}.
\newblock Phd thesis, Massachussetts Institute of Technolgy, Cambridge,MA, september 1961.

\bibitem{simon98}
R.~Simon and N.~Mukunda.
\newblock Twist phase in gaussian-beam optics.
\newblock {\em J. Opt. Soc. Am. A}, 15(9):2373, 1998.

\bibitem{Cai17}
Yangjian Cai, Yahong Chen, Jiayi Yu, Xianlong Liu, and Lin Liu.
\newblock Chapter three - generation of partially coherent beams.
\newblock volume~62 of {\em Progress in Optics}, pages 157--223. Elsevier, 2017.

\bibitem{wang20}
Rui Wang, Shijun Zhu, Yikai Chen, Hongkun Huang, Zhenhua Li, and Yangjian Cai.
\newblock Experimental synthesis of partially coherent sources.
\newblock {\em Opt. Lett.}, 45(7):1874--1877, Apr 2020.

\bibitem{tian20}
Cong Tian, Shijun Zhu, Hongkun Huang, Yangjian Cai, and Zhenhua Li.
\newblock Customizing twisted schell-model beams.
\newblock {\em Opt. Lett.}, 45(20):5880--5883, Oct 2020.

\bibitem{wang10}
Fei Wang and Yangjian Cai.
\newblock Second-order statistics of a twisted gaussian schell-model beam in turbulent atmosphere.
\newblock {\em Opt. Express}, 18(24):24661--24672, Nov 2010.

\bibitem{liu19}
Yonglei Liu, Xianlong Liu, Lin Liu, Fei Wang, Yuping Zhang, and Yangjian Cai.
\newblock Ghost imaging with a partially coherent beam carrying twist phase in a turbulent ocean: A numerical approach.
\newblock {\em Applied Sciences}, 9(15), 2019.

\bibitem{zhou20}
Mengyao Zhou, Weichen Fan, and Gaofeng Wu.
\newblock Evolution properties of the orbital angular momentum spectrum of twisted gaussian schell-model beams in turbulent atmosphere.
\newblock {\em J. Opt. Soc. Am. A}, 37(1):142--148, Jan 2020.

\bibitem{Hutter20}
Lucas Hutter, G.~Lima, and S.~P. Walborn.
\newblock Boosting entanglement generation in down-conversion with incoherent illumination.
\newblock {\em Phys. Rev. Lett.}, 125:193602, Nov 2020.

\bibitem{Hutter21}
Lucas Hutter, E.~S. Gomez, G.~Lima, and S.~P. Walborn.
\newblock Partially coherent spontaneous parametric downconversion: Twisted gaussian biphotons.
\newblock {\em AVS Quantum Sci.}, 3:031401, 2021.

\bibitem{ismail2020}
Samukelisiwe~Purity Phehlukwayo, Marie~Louise Umuhire, Yaseera Ismail, Stuti Joshi, and Francesco Petruccione.
\newblock Influence of coincidence detection of a biphoton state through free-space atmospheric turbulence using a partially spatially coherent pump.
\newblock {\em Phys. Rev. A}, 102:033732, Sep 2020.

\bibitem{Santos2022}
Gustavo~H dos Santos, Andre~G de~Oliveira, Nara~Rubiano da~Silva, Gustavo Ca{\~n}as, Esteban~S G{\'o}mez, Stuti Joshi, Yaseera Ismail, Paulo H~Souto Ribeiro, and Stephen~Patrick Walborn.
\newblock Phase conjugation of twisted gaussian schell model beams in stimulated down-conversion.
\newblock {\em Nanophotonics}, 11(4):763--770, 2022.

\bibitem{mair01}
Alois Mair, Alipasha Vaziri, Gregor Weihs, and Anton Zeilinger.
\newblock Entanglement of the orbital angular momentum states of photons.
\newblock {\em Nature}, 412:313, 2001.

\bibitem{caetano02}
D.~P. Caetano, M.~P. Almeida, P.~H. Souto~Ribeiro, J.~A.~O. Huguenin, B.~Coutinho~dos Santos, and A.~Z. Khoury.
\newblock Conservation of orbital angular momentum in stimulated down-conversion.
\newblock {\em Phys. Rev. A}, 66(4):041801, 2002.

\bibitem{walborn04a}
S.~P. Walborn, A.~N. de~Oliveira, R.~S. Thebaldi, and C.~H. Monken.
\newblock Entanglement and conservation of orbital angular momentum in spontaneous parametric down-conversion.
\newblock {\em Phys. Rev. A}, 69:023811, 2004.

\bibitem{Pereira17}
Leonardo~J. Pereira, Wagner~T. Buono, Daniel~S. Tasca, Kaled Dechoum, and Antonio~Z. Khoury.
\newblock Orbital-angular-momentum mixing in type-ii second-harmonic generation.
\newblock {\em Phys. Rev. A}, 96:053856, Nov 2017.

\bibitem{Oliveira21}
A.G. de~Oliveira, G.~Santos, N.~Rubiano da~Silva, L.J. Pereira, G.B. Alves, A.Z. Khoury, and P.H.~Souto Ribeiro.
\newblock Beyond conservation of orbital angular momentum in stimulated parametric down-conversion.
\newblock {\em Phys. Rev. Appl.}, 16:044019, Oct 2021.

\bibitem{canas22}
G~Cañas, E~S Gómez, G~H dos Santos, A~G de~Oliveira, N~Rubiano da~Silva, Stuti Joshi, Yaseera Ismail, P~H~S Ribeiro, and S~P Walborn.
\newblock Evaluation of twisted gaussian schell model beams produced with phase randomized coherent fields.
\newblock {\em J. Opt.}, 24(9):094004, 2022.

\bibitem{deOliveira24}
M.~Gil {de Oliveira}, A.L.S. Santos, A.C. Barbosa, B.~Pinheiro {da Silva}, G.H. {dos Santos}, G.~Cañas, P.H.~Souto Ribeiro, S.P. Walborn, and A.Z. Khoury.
\newblock Anomalous second harmonic generation of twisted gaussian schell model beams.
\newblock {\em Optics \& Laser Technology}, 176:110983, 2024.

\end{thebibliography}


\end{document}